\documentclass[aps,pre,twocolumn,showpacs]{revtex4-1}  
\usepackage[dvipsnames]{xcolor}
\usepackage{graphicx}  
\usepackage{dcolumn}   
\usepackage{bm}       
\usepackage{amssymb}  
\usepackage{amsmath}
\usepackage{float}
\usepackage{enumerate}
\usepackage{braket}
\begin{document}
\title{Stable Bloch oscillations and Landau-Zener tunneling in a non-Hermitian $\cal{PT}$-symmetric flat band lattice}
\author{J. Ramya Parkavi$^1$, V. K. Chandrasekar$^1$ and M. Lakshmanan$^2$ }
\address{$^1$ Centre for Nonlinear Science \& Engineering, School of Electrical \& Electronics Engineering, SASTRA Deemed University, Thanjavur-613 401, Tamilnadu, India.\\
$^2$ Centre for Nonlinear Dynamics, School of Physics, Bharathidasan University, Tiruchirappalli 620 024, Tamilnadu, India.}
 \begin{abstract}
 	\par This article aims to study the existence of stable Bloch oscillations and Landau-Zener tunneling in a non-Hermitian system when exposed to external fields. We investigate a non-Hermitian $\cal{PT}$-symmetric diamond chain network and its transport dynamics in two different situations, namely in a flat band case and a non-flat band case. The considered system does not support unbroken-$\cal{PT}$ phase or completely real eigenspectra in any of the parametric regions in both the flat and non-flat band cases. In the flat band case, up to a critical value of the gain-loss parameter, the bands are found to be gapless or inseparable, and for other values the bands are isolated. Considering the non-flat band case, all the bands are found to be complex dispersive and are also isolated. In the case of completely broken $\cal{PT}$ phase, we look upon the possibility to have stable dynamics or Bloch oscillations upon the application of external fields like synthetic electric field. In particular, when the complex bands are isolated, we point out that the Landau-Zener tunneling induced by the synthetic electric field can enable Bloch oscillations. The amplitude of these Bloch oscillations is large and persists for a long propagation distance which reveals that super Bloch oscillations can be observed in the broken $\cal{PT}$ phase of the system. We also report the amplified Bloch oscillations which pave the way towards controlling transport phenomena in non-Hermitian systems.
\end{abstract}
\maketitle
\section{Introduction}
\par Understanding the dynamics of a quantum system in a periodic potential has been an active research field for the past few decades. To describe the dynamics of a particle in the potential and to localize the transport in a periodic structure, two underlying phenomena viz Bloch oscillations and Landau-Zener tunneling are being investigated. When the quantum particle in the periodic potential is driven by an external field, it performs a localized oscillatory motion which is termed as Bloch oscillations  \cite{blochorg, blochzener}. If the applied external field is sufficiently strong, tunneling between the bands emerges which is popularly known as Landau-Zener (LZ) tunneling \cite{landau,landauzener}. Initially, Bloch oscillations were observed in semiconductor superlattices  \cite{semiconductor, submilli}. Later on, Bloch oscillations were shown to occur in different periodic systems such as ultracold atoms \cite{ultracold, ultracold2},
Bose-Einstein condensates in optical lattices \cite{bec, bec2},
 waveguide arrays \cite{wavegd1, wavegd2, wavegd3},
optically induced lattices \cite{opticlatc, opticlatc2},
acoustical waves \cite{acoustic}
and plasmonic systems \cite{plas}. 
Among the various physical settings, optical systems offer direct visualization of Bloch oscillations \cite{ultracold, wavegd1, bec2, blochopt}
and Landau-Zener tunneling \cite{lzopt,lzopt2,lzopt3,lzopt4}
both theoretically and experimentally. 
For example, Bloch oscillations with Landau-Zener tunneling can be used to construct matter-wave beam splitters and Mach-Zehender interferometer \cite{machzen}. 
 In this context, the interplay of Bloch oscillations and Landau-Zener tunneling has attracted wide research interest \cite{opticlatc, blochzen, blochzen3}.
\par Further, differing from the symmetric Bloch oscillations and Landau-Zener tunneling, asymmetric cases were also reported in one dimensional tight binding model and Bose-Einstein condensates in an optical lattice with nonlinear interactions \cite{ultracold, dynamicsbo, lzbec1, lzbec2}. More recently, in Ref. \cite{flat}, the authors have considered a Hermitian three leg diamond network and reported asymmetric Bloch oscillations with Landau-Zener tunneling due to the presence of flat band. Similarly, the authors of Ref. \cite{spinorbit} have considered optical lattices with spin orbit coupling and shown the suppression of Bloch oscillations due to the flattening of bands. From the non-Hermitian perspective, in Ref. \cite{nonhermcoup}, the authors have shown the revival of localized oscillations (Aharanov-Bohm cages) and  unstable amplification of Bloch waves due to the complex flat band nature induced by non-Hermitian coupling in the system. The earliest study of flat band was carried out in a  $\cal{PT}$-symmetric diamond chain network \cite{compacton} by Yulin and Konotop and further the works in refs. \cite{nonhermflat,flatinterplay} also emphasize the beneficial impact of flat bands in the localization mechanism of non-Hermitian systems. On the other hand, Bloch oscillations and Landau-Zener tunneling were well-identified in two layers or systems (both Hermitian and non-Hermitian) with two dispersive bands \cite{machzen, nonlinbo, twobands,  selfimage, twobandspt}. More exploration of these studies were needed in the system involving three bands, particularly flat bands. Attracted by these studies, we are here interested to explore the non-Hermitian version of asymmetric/stable Bloch oscillations with Landau-Zener tunneling in the trilayer flat band lattices under the effect of external fields. 
\par Recently, several intriguing features of non-Hermitian $\cal{PT}$-symmetric systems such as unidirectional invisibility \cite{invisb, invisb2, invisb3}, simultaneous lasing-absorbing \cite{lasabs, lasabs2}
and selective mode lasing \cite{modlas, modlas2} have triggered interest in the study of Bloch oscillations and Landau-Zener tunneling in $\cal{PT}$-symmetric systems. The first theoretical investigation of Bloch oscillations in a $\cal{PT}$-symmetric complex crystal was addressed in \cite{unibloch2}, where it was shown that amplified/attenuated oscillations can take place depending on the nature of the force. Experimental realization of Bloch oscillations was also demonstrated in $\cal{PT}$-symmetric global and local mesh of lattices \cite{global}. Especially, Bloch oscillations in $\cal{PT}$-synthetic photonic lattice expands the applicability of on-chip photonics \cite{blochexpr}. Likewise, Landau-Zener transitions were investigated in a $\cal{PT}$-symmetric optical lattice \cite{blochexpr2}, which is useful to control the intensity of light beam in complex waveguide arrays. In \cite{selfimage}, the authors have shown the possibility of stable Bloch oscillations with Zener tunneling at the exact $\cal{PT}$ phase of the photonic lattices. Also, the observed oscillations experienced a wave-packet self-imaging and giant recombinations of beams which are useful for the realization of beam splitters and image processing. Interestingly, in the $\cal{PT}$ phase of the $\cal{PT}$-symmetric system \cite{super}, one can observe large amplitude matter-wave oscillations referred as super Bloch oscillations. Although, the super Bloch oscillations have been explored in the $\cal{PT}$ phase of the system, achieving this type of large amplitude Bloch oscillations in the broken $\cal{PT}$ phase is a challenging problem and still remains to be investigated.  In the course of this present work, we also try to exploit the large amplitude Bloch oscillations in the broken regime of the $\cal{PT}$-symmetric system.
\par Focusing on the recent interest over the Bloch oscillations with Landau-Zener tunneling in the non-Hermitian systems and the systems supporting non-dispersive flat bands, we here consider the light transport in a non-Hermitian diamond chain lattice where the system is found to support a flat band in a particular situation. By considering the flat band and non-flat band cases, we study the dynamics and possibilities of localization upon the application of external fields like synthetic electric and magnetic fields. Apart from being non-Hermitian, the system is also found to be $\cal{PT}$-symmetric in particular situations (that is in the absence of transverse component of the electric field). However, the system does not support complete real eigenspectra in any of the parametric regions and the $\cal{PT}$-symmetry is found to be spontaneously broken throughout the parametric space. Thus, the question that we put forth here is, ``Is it possible to have stable dynamics in this system (supporting only broken phase) through the application of external fields?" Particularly, we study the impact of externally applied field in the situations having gapless and isolated band structures and also in situations involving  flat band and non-flat band cases. In the case of complex isolated band structure, we study whether the Landau-Zener transitions among the gain and loss bands (bands with $Im(\lambda)>0$ and $Im(\lambda)<0$ respectively, where $\lambda$ is the eigenvalue) can make stable Bloch oscillations possible. 
  \par To explore the above, this article is structured in the following manner.  In Sec. II, we present the non-Hermitian model under consideration and investigate the situations corresponding to flat band and non-flat band cases. Sec.  III explains the transport dynamics of the lattice in the flat band case where we show the existence of compact localized modes supported by the flat band in the absence of external fields and the possibility of Bloch oscillations  with respect to the applied field.  Similarly, the lattice dynamics in the non-flat band case is discussed in the Sec. IV. Finally, a summary of the obtained results is given in Sec. V. 
 \section{\label{mod}Model}
\begin{figure}[h]    
 \includegraphics[width=1.0\linewidth]{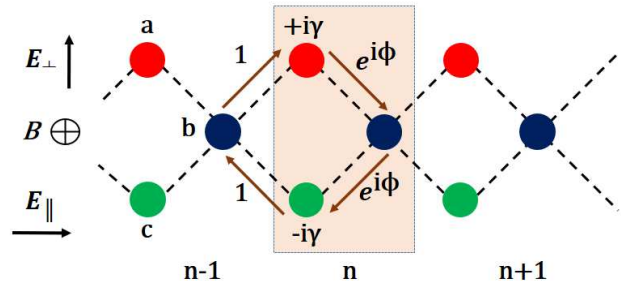}
 \caption{The n$^{th}$ unit cell of an array of non-Hermitian lattice composed of an amplifying waveguide (at site a), a passive waveguide with no loss or gain (at site b) and a dissipating waveguide (at site c). Single-mode channel waveguides are placed at each site in such a way that the modes overlap. Dashed lines indicate the sites connected with the hopping of light. Solid arrows indicate the phase of complex hopping constants $\phi$  for the specific synthetic magnetic field $B$. $E_\parallel$ and $E_\perp$ define the synthetic electric fields along the longitudinal and transversal directions, respectively.}
 			\label{fig0}
\end{figure}

We consider a non-Hermitian diamond chain lattice which is made up of an array of waveguides. The schematic diagram of the lattice model under consideration is given in Fig. \ref{fig0}. The system has three layers, namely $a$, $b$ and $c$-layers. The sites of top and bottom layers ($a$ and $c$ layers) of the lattice have a gain and loss nature, respectively, and thus make the system to be non-Hermitian and also $\cal{PT}$-symmetric. To study the transport dynamics of light in the presence of external fields, the system is subject to synthetic electric and magnetic fields. Originally, this type of $\cal{PT}$-symmetric diamond chain model was introduced in \cite{compacton} in the absence of an electric field and in the later work, Hermitian version of similar model was studied with the role of electric field components \cite{flat} inlcuded.

The dynamics of the evolving electric field amplitude at the $n^{th}$ unitcell, $\psi_{n}(z)=(a_{n}(z), b_{n}(z), c_{n}(z))^{T}$ is given by the following equations,
\begin{eqnarray}
&&i\dot{a_n} = (E_{\parallel}n+E_{\perp})a_n+i\gamma a_n-e^{-i\phi} b_n-b_{n-1}, \nonumber \\
&&i\dot{b_n} = E_{\parallel}(n+\frac{1}{2})b_n-e^{i\phi}a_n-e^{-i\phi}c_n-c_{n+1}-a_{n+1}, \nonumber \\
&&i\dot{c_n} = (E_{\parallel}n-E_{\perp})c_n-i\gamma c_n-e^{i\phi}b_n-b_{n-1}.
\label{eq2a}
\end{eqnarray}
In the above Eq. (\ref{eq2a}) ${a}_{n}$, ${b}_{n}$ and ${c}_{n}$ are the complex field amplitudes in the waveguides and the overdot represents differentiation with respect to $z$, where $z$ represents the propagation distance in dimensionless unit. The gain and loss parameter ($\gamma$) can be introduced through a proper choice of the complex refractive index profiles  \cite{ptop, ptop2}, particularly by tuning the imaginary part of the refractive index. The waveguides in the layer $a$ are chosen to be of amplifying type and the ones in the layer $c$ are chosen to be of dissipative type. $E_\parallel$ and $E_\perp$ respectively denote the components of the synthetic electric field along the longitudinal and transverse directions to the lattice plane. $E_\parallel$ and $E_\perp$ (simply termed here as longitudinal and transverse electric fields, respectively, throughout the manuscript) can be realized by modulating the refractive index gradient \cite{modulelec,modulelec2}, for example by applying a temperature gradient across the thermo-optical waveguide material \cite{blochopt, tempert}, or by a circular bending of the waveguides \cite{bending1, bending2, geometry}, or by a proper choice of the waveguide geometries \cite{wavegd1, bending3}. In the context of optical or photonic lattices,  the synthetic magnetic field B can be engineered by direction dependent phase factor $e^{\pm i\phi}$ introduced through optical path imbalance $\Delta x$ in the tunneling between primary and auxiliary resonators (or waveguides), resulting in the magnetic flux  $\phi=\frac{2\pi \Delta x}{\lambda_{r}}$ ($\lambda_{r}$ is the resonant wavelength) \cite{opticalimbalance3, opticalimbalance2, opticalimbalance1}. In the periodically driven optical lattices the value of $\phi$ is restricted to either $0$ or $\pi$ as in Refs. \cite{mag0phi1, mag0phi2, mag0phi3}. Also, by suitable driving, the phase can be tuned to any value $\phi \in [0,2\pi)$ in the optical lattices as mentioned in \cite{mag02pi, tunablegauge}. Alternate methods to generate the artificial magnetic field was suggested by Fang et al. in \cite{magfield} and it can be realized by sinusoidal modulation of the refractive index of the waveguides \cite{abcage2} or by proper longitudinal modulation of the propagation constant of the waveguides  \cite{magfield2} or by specially fabricated waveguides and its surrounding media \cite{prlmagfield3}.
\par In the framework of atomic physics, one can construct the equivalent non-Hermitian Hamiltonian of the present model via anihilation and creation operators, leading to an equation similar to (\ref{eq2a}) for the probability amplitudes. The considered model may be realized in the theme of atomic settings by defining the parameter $\gamma$ as the gain-loss strength which determines the level of non-Hermiticity. The non-Hermitian $\cal{PT}$-symmetric nature for the $N$-site of the system can be incorporated by combining an absorbing potential on one site with an emitting potential on another site  as given in \cite{Nsitenonherm1, Nsitenonherm2, Nsitenonherm3, Nonhermscatter}. In the presence of $\gamma$, the system  given in Eq. (\ref{eq2a}) can be considered as the non-Hermitian extension of the model considered in \cite{flat}. $E_\parallel$ and $E_\perp$ respectively denote the components of the electric field along the longitudinal and transversal directions to the lattice plane. In the presence of a magnetic field $B$, the hopping terms between the neighbouring pairs $i$ and $j$ acquire additional phase factors $e^{i\phi_{ij}}$ involving the vector potential ${\bf A}$: $\phi_{ij}=\frac{2\pi}{\Phi_{0}} \int_{i}^{j} {\bf A}.d{\bf l}$, where $\Phi_{0}=\frac{hc}{e}$ is the flux quantum. For the considered case, the whole spectrum depends only on the reduced flux $f=\frac{\Phi}{\Phi_{0}}$, where $\Phi=\frac{Bq^2}{2}$ is the magnetic flux through an elementary diamond ($q$ is the unit cell vector length). The phase $\phi=2\pi f (\phi \in [0,2\pi))$ makes the tunneling amplitude to be complex and it can be introduced through a proper dc magnetic field $B$ (generated by artificial gauge field) that is oriented perpendicular to the plane embedding the diamond lattice chain \cite{magfluxdia, abcage2d, abcage}. 
\par  The considered system as given in Eq. $(\ref{eq2a})$ is $\cal{PT}$-symmetric in the absence of transverse electric field where the system is invariant under the combined operation of parity and time-reversal symmetries defined by $a_{n}\rightarrow-c_{n}$, $b_{n}\rightarrow-b_{n}$, $c_{n}\rightarrow-a_{n}$, $i\rightarrow-i$ and $z \rightarrow -z$.

\par In the absence of longitudinal electric field (i.e., $E_\parallel=0$), the eigenmodes can be written as $\{a_n,b_n,c_n\}=(A,B,C)$exp$(i\lambda z + ikn)$, where $\lambda$ denotes the propagation constant and $k \in \cal{R}$ denotes the Bloch wave vector. After the substitution of the above form into Eq. (\ref{eq2a}) we obtain the following characteristic equation:
\begin{equation}
\lambda^{3}-P\lambda+Q=0,
\label{eq4} \nonumber
\end{equation}
where
\begin{eqnarray}
P&=&E_\perp^{2}+2i\gamma E_\perp-\gamma^{2}+4(1+cos \phi \cos k), \nonumber\\
Q&=&4 E_\perp \sin \phi \sin k+4 i \gamma \sin \phi \sin k.
\label{eq4a}
\end{eqnarray}
\par From Eq. (\ref{eq4a}), it is clear that whenever $Q=0$, one of the propagation constants (eigenvalues) becomes zero, that is, in the case of $\phi=l \pi$, $l=0,1$. The existence of zero propagation constant for $\phi=l \pi$ indicates the presence of the flat band and so we explore the dynamics in two different situations, namely (i) in the flat band case ($\phi=l \pi$, $l=0,1$.) and (ii) in the non-flat band case ($\phi \neq l\pi$).
\section{Dynamics in the flat band case}
In this section we present the transport dynamics of light in the presence and absence of external fields.
\subsection{\label {eigenbroken} In the absence of electric field components}
\begin{figure}[h]
	\begin{center} 
		\includegraphics[width=1.0\linewidth]{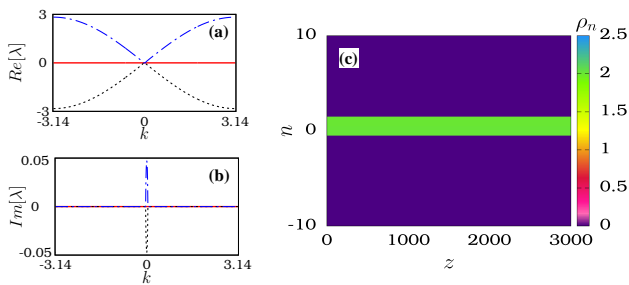}
	\end{center}
	\caption{(Color online) (a) and (b) Real and imaginary parts of the band structure for a $\cal{PT}$-symmetric system with phase $\phi=\pi$ and in the absence of synthetic electric field components ($E_\parallel=0$, $E_\perp=0$). Gain-loss parameter is chosen as $\gamma=0.05$. The solid red line represents the non-dispersive (flat) band. Dashed black and dash-dotted blue curve represents the complex dispersive nature of the bands. (c) Localization of light for the CLS type initial condition as given in Eq. (\ref{eq7s}).}
	\label{fig2}
\end{figure}

Firstly, we consider the situation in which the electric field components are absent and then present the compact localized modes supported by the flat band of the system. As mentioned earlier, the flat band arises in the case of $\phi=l \pi$, $l \in {0, 1}$ (that is, the situation in which there is no complex hopping). We recall here that the criterion for the existence of a flat band in the Hermitian case is the same as that of the non-Hermitian case. However, in contrast to the Hermitian case, here the flat band exists along with complex dispersive bands rather than with real dispersive bands. For instance, we present below the eigenspectra corresponding to two different situations, namely $\phi=0$ and $\phi=\pi$.
\begin{small}
	\begin{eqnarray}
	\phi&=&0:\quad\lambda_1=0,\quad \lambda_{2,3}=\mp\sqrt{-\gamma^{2}+4(1+\cos k)},\label{eq5}\\
\phi&=&\pi:\quad \lambda_1=0,\quad \lambda_{2,3}=\mp\sqrt{-\gamma^{2}+4(1-\cos k)}.
\label{eq5a}
 \end{eqnarray}
 \end{small}
From the above Eqs. (\ref{eq5}) and (\ref{eq5a}), the eigenvalue $\lambda_{1}$ implies the non-dispersive flat band (as it is independent of $k$) and the other two eigenvalues $\lambda_{2}$ and $\lambda_{3}$ indicate the dispersive bands. It is obvious from the eigenvalues that complete real eigenspectra are not possible for any parametric value and so the $\cal{PT}$-symmetry is spontaneously broken for all parametric values. Secondly, considering the region $\gamma \leq \gamma_{c}=2 \sqrt{2}$, the three bands meet together in both the cases, $\phi=0$ and $\phi=\pi$. Thus, the bands are found to be gapless or inseparable \cite{PTbands}. However, considering the regime $\gamma > \gamma_{c}=2 \sqrt{2}$,  the complex energy bands are found to be isolated (or gapped) \cite{PTbands} where 
$\lambda_n(k) \neq \lambda_m(k^\prime)$ for all $k, k^\prime \in [-\pi,\pi]$. The band structure for 
$\gamma\leq\gamma_{c}$ is given in Figs. \ref{fig2}(a) and \ref{fig2}(b) for the case $\phi=\pi$. The figures show the existence of a pure real flat band between the complex dispersive bands. One of the dispersive bands has $Im[\lambda]>0$ indicating amplifying nature and the other has $Im[\lambda]<0$ indicating dissipative nature.
\par The pure real flat band allows the lattice to support exact eigenmodes in the form of compact localized eigenstates (CLSs) which include non-zero amplitudes at a finite number of sites and vanishing amplitudes at all other sites. The compact localized eigenstates for a non-Hermitian $\cal{PT}$-symmetric diamond chain lattice were first obtained in \cite{compacton} and they are popularly known as compactons. For instance, in our considered model in the case of $\phi=0$, the eigenmodes corresponding to the flat band satisfy the relations, namely, $A_{n}=- C_{n}$ and $B_{n}+B_{n-1}=i \gamma A_{0}$, $n=-N,-(N-1),..,0,..,N$. Due to this, the system can admit finite site localization and we provide possible forms of compact localization modes below,
\begin{small}
\begin{eqnarray}
A_{n}&=& -C_{n}=A_0 \delta_{0,n}, \nonumber \\
B_{n}&=&(-1)^{n} i \gamma A_{0}\delta_{m,n}, \quad m=0,1,2,3...
\end{eqnarray}
\end{small}\\
In the above, even though localization to a single site (here localization at site zero is demonstrated but the localization can be achieved at any site) has been achieved in the layers $a$ and $c$, such type of finite site localization is not achieved in layer $b$. Instead, if we consider two site localization in layers $a$ and $c$, finite site localization can be seen in the layer $b$ also where the CLS mode can take the form
\begin{small}
\begin{eqnarray}
A_{n}&=&-C_{n}=A_{0} \delta_{s,n}, \quad where \quad s=0 \quad and \quad 1, \nonumber \\
B_{n}&=&i\gamma A_{0} \delta_{0,n}.
\end{eqnarray}
\end{small}
Similarly, in the case of $\phi=\pi$, the CLS takes the form 
\begin{small}
\begin{eqnarray}
A_{n}&=&-C_{n}=(-1)^{n}A_{0} \delta_{s,n},  \quad where \quad s=0 \quad and \quad 1, \nonumber\\
B_{n}&=&-i\gamma A_{0} \delta_{0,n}.
\label{eq7s}
\end{eqnarray}
\end{small}
Thus for the CLS type initial configuration, the system enables localization into finite sites of the lattice. For instance, by considering the initial condition in the form of CLS (as given in Eq. (\ref{eq7s})) and choosing $A_0=1$, we have plotted the intensity, $\rho_{n}=(|A_n|^2+|B_n|^2+|C_n|^2)$, evolution in the case of $\phi=\pi$. Actually, we have considered $n=-150,-149,..,0,..,150$ unit cells and for the clear visualization of localization we have shown Fig. 2(c) only for $n=-10,-9,..,0,..,10$ unit cells. The figure clearly demonstrates the localization into finite number of sites. Due to this fact, the finite site localization can be achieved for any value of $\gamma$ for CLS initial condition in the absence of external fields $E_\parallel$ and $E_\perp$.
\par The question that arises next is the behaviour of the flat band lattice in the presence of electric fields and it will be seen in the following. 
 \subsection{\label{ptbo}Asymmetric Bloch oscillations in the presence of $E_\parallel$}
Before considering both the longitudinal and transverse electric fields ($E_\parallel$ and $E_\perp$ together), we first study the role of $E_\parallel$. As discussed in Sec. \ref{mod}, adding longitudinal electric field still preserves the $\cal{PT}$-symmetric nature of the system. We also recall here that in the absence of $E_\parallel$ alone, the band structure of the system has zero band gap in the region of $\gamma \leq \gamma_{c}$ and the bands are isolated in the region of $\gamma > \gamma_{c}$.
\begin{figure}[h]
	\centering
	\includegraphics[width=1.025\linewidth]{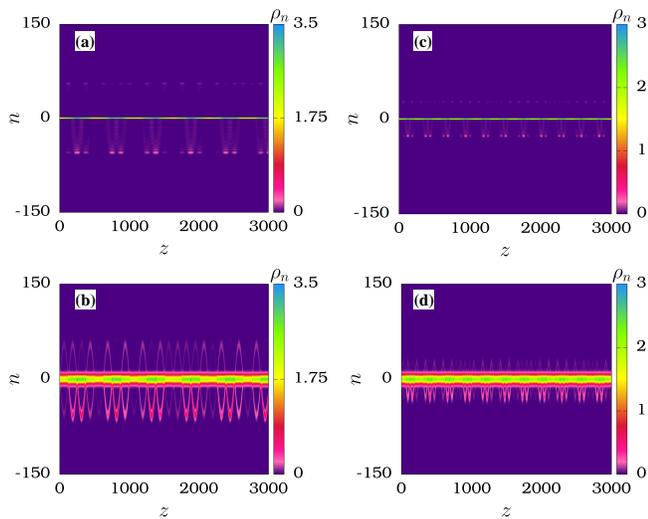}
	\caption{(Colour online) Numerical simulations showing asymmetric Bloch oscillations for the CLS initial excitation (top row) and broad initial Gaussian excitation (bottom row) in the presence of  magnetic field with phase $\phi=\pi$. The figures are plotted for different values of longitudinal field, namely $E_\parallel=0.05$ in (a), (b) , and $E_\parallel=0.1$ in (c), (d) with the gain-loss strength $\gamma=0.05$.}
		\label{fig3}
\end{figure}
\par To elucidate the transport dynamics in the presence of longitudinal electric field, we consider the flat band case $\phi=\pi$ in the region $\gamma\leq\gamma_{c}$ and obtain the evolution patterns for two different initial excitations, namely, (i) CLS type initial condition as given in Eq. (\ref{eq7s}) and (ii) Broad Gaussian type excitation given by $A_{n}(0)=-C_{n}(0)=e^{-\frac{n^2}{2 \sigma^2}}$, where $\sigma=70$ and $B_{n}(0)=0$. A very small value of longitudinal electric field, say $E_\parallel=0.05$, is enough to simulate asymmetric Bloch oscillations in the system corresponding to both types of initial excitations as shown in Figs. \ref{fig3}(a) and \ref{fig3}(b). From Figs. \ref{fig3}(c) and \ref{fig3}(d), well localized asymmetric Bloch oscillations at the nearest sites are observed for the value of $E_\parallel=0.1$. This shows that strong localization to a few sites in the form of Bloch oscillations requires a higher longitudinal electric field.
\par The results observed in Figs. \ref{fig3}(a)-\ref{fig3}(d) also show that the observed Bloch oscillations are asymmetric with respect to the $n=0$ site. Also, the pattern is not periodic with respect to $z$ and depicts asymmetric evolution. We also recall here that even in the Hermitian lattices, this type of asymmetric Bloch oscillations are not usual and they are observed only in the lattices supporting flat bands \cite{flat}.
 \par As discussed in Sec. \ref{eigenbroken}, in the absence of $E_\parallel$, the real flat and complex dispersive bands are found to be inseperable in the region of $\gamma\leq\gamma_{c}$ and the complex nature of these dispersive bands leads to the $\cal{PT}$-symmetric broken nature of the system. Likewise, in the presence of $E_\parallel$ also,  the dispersive bands are found to be complex which retains the broken $\cal{PT}$-symmetric nature of the system. On the otherside, as soon as the longitudinal electric field is introduced to the system, the existence of real flat band between the complex dispersive bands (one band has $Im[\lambda]<0$ and the other band has $Im[\lambda]>0$) in the gapless situation leads to the interaction of CLS sites with its neighbouring sites. As a result, compact localized initial state starts to evolve in an asymmetric manner in the form of stable Bloch oscillations. From the results, we conclude that even though the  $\cal{PT}$-symmetry is broken here, the observed Bloch oscillations neither get amplified nor attenuated with respect to propagation distance $z$ in the presence of longitudinal electric field.  However, these stable Bloch oscillations are not observed in the region of $\gamma>\gamma_{c}$ where the bands are isolated. In the latter region, we observe blow-up type responses only.
\subsection{Amplifying Bloch oscillations: in the presence of both $E_\parallel$ and $E_\perp$}
\begin{figure}[h]
	\centering
	\includegraphics[width=0.975\linewidth]{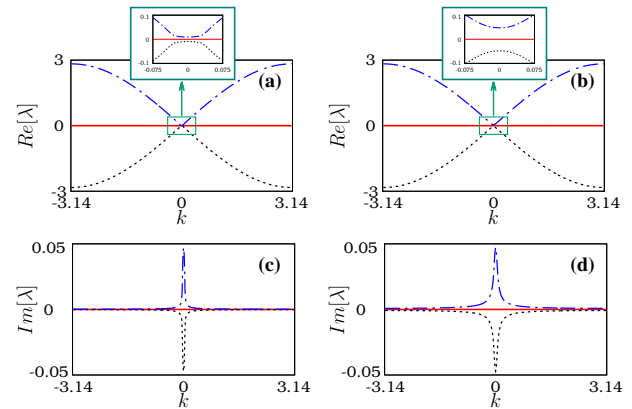}\\
	\caption{(Colour online) (a), (b) Real parts and (c), (d) Imaginary parts of the band structures in the presence of transverse electric fields ($E_\perp$). (a), (c) for $E_\perp=0.01$, (b), (d) for $E_\perp=0.05$. The upper insets show the emergence of bandgaps at the center of the band structures. Other parameters: $\gamma=0.05$, $\phi=\pi$ and $E_\parallel=0$.}
		\label{fig4}
\end{figure}

In the previous section, we have shown the existence of stable Bloch oscillations (in the region $\gamma \leq \gamma_{c}$) in the absence of transverse electric field.  Now, the introduction of $E_\perp$ destroys the $\cal{PT}$-symmetric nature of the system ($\cal{PT}$-symmetry is explicitly broken).  So, the real eigenspectra may not be possible in any of the parametric regions.  For instance, the eigenspectra of the flat band case in the presence of $E_{\perp}$ (in the absence of $E_\parallel$) can be given as follows:\\
\begin{small}
(i): $\phi=0$\\
 \vspace{-0.45cm}
	\begin{eqnarray}
 \lambda_1&=&0,\quad \lambda_{2,3}=\mp\sqrt{E_{\perp}^{2}+2i\gamma E_{\perp}-\gamma^2+4(1+\cos k)}. \quad
 \end{eqnarray}
(ii): $\phi=\pi$\\
 \vspace{-0.45cm}
 \begin{eqnarray}
\lambda_1&=&0,\quad \lambda_{2,3}=\mp\sqrt{E_{\perp}^{2}+2i\gamma E_\perp-\gamma^2+4(1-\cos k)}. \quad
	\label{eqa5}
	\end{eqnarray}
\end{small}
It is obvious from the above equations that the eigenvalues $\lambda_{2,3}$ are complex in all the parametric regions. The band structure in the case of $\phi=\pi$ is presented for two different values of $E_\perp$ in Figs. \ref{fig4}(a)-\ref{fig4}(d). It is also obvious from Figs. \ref{fig4}(a)-\ref{fig4}(d) that in both the cases of $E_\perp=0.01$ and $E_\perp=0.05$, the bands are isolated and the bandgap is widened with the increase of $E_\perp$. 
\par Even in this case with $E_\perp$, the flat bands support compact localized modes which can be given by\\
For $\phi=0$:
 \vspace{-0.175cm}
\begin{eqnarray}
A_{n}&=&-C_{n}=A_{0} \delta_{s,n} \quad where \quad s=0 \quad and \quad 1 \nonumber \\
B_{n}&=&(E_{\perp}+i\gamma) A_{0} \delta_{0,n}.
\label{eq6a}
\end{eqnarray}
For $\phi=\pi$:
 \vspace{-0.175cm}
\begin{eqnarray}
A_{n}&=&-C_{n}=(-1)^{n}A_{0} \delta_{s,n}  \quad where \quad s=0 \quad and \quad 1 \nonumber\\
B_{n}&=&(-E_{\perp}-i\gamma) A_{0} \delta_{0,n}.
 \label{eq71}
 \end{eqnarray}
\par
\begin{figure}[h]
	\centering
	\includegraphics[width=1.0\linewidth]{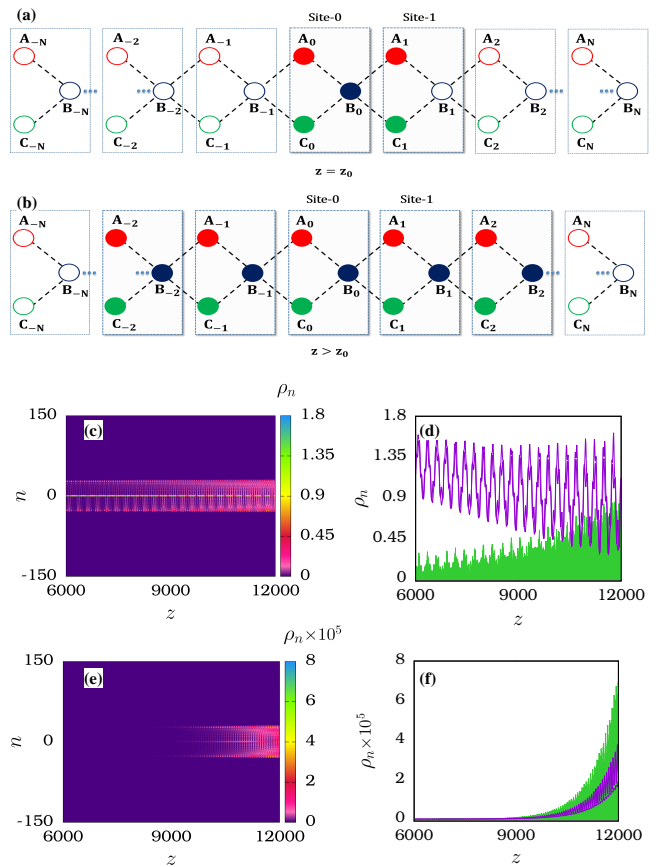}\\
	\caption{(Colour online) (a) CLS modes initially excited at site-0 and site-1 as given in Eq. (\ref{eq71}) at $z=z_{0}$ ($z_{0}=0$). (b) depicts the change of CLS modes and give rise to oscillation pattern in $z>z_{0}$ due to Landau-Zener tunneling when $E_\parallel$ is turned on to be nonzero. Filled circles denote the sites with non-zero amplitude and empty circles represent the sites with zero amplitude. (c) and (e) shows amplified Bloch oscillations for two different values of transverse electric field after long transients. (d) and (f) Evolution of intensity ($\rho_{n}$) with respect to propagation distance $z$. The solid violet curve represents the intensity in the initially excited CLS sites ($A_{0}, B_{0}, C_{0}, A_{1}$ and $C_{1}$) and the solid green curve represents the intensity in the remaining sites (i.e. initially unexcited sites). The parameter values in (c), (d) $E_\perp=0.01$, (e), (f) $E_\perp=0.05$ with $E_\parallel=0.1$, $\phi=\pi$, and $\gamma=0.05$.}
		\label{fig5}
\end{figure}

  Now, the inclusion of the longitudinal electric field $E_\parallel$ may induce Landau-Zener transition among the complex dispersive bands and the real flat band. As a result, Bloch oscillations emerges in the system. In order to find the transport dynamics with the application of $E_\parallel$, we first consider the CLS type initial condition given in Eq. (\ref{eq71}) and find how the CLS evolution is perturbed by $E_\parallel$. As we have considered CLS type initial condition in these cases, the sites $0$ and $1$ alone are excited initially as shown in Fig. \ref{fig5}(a). As $z$ increases, compact localized states occupy more sites than the CLS sites and amplification occurs because of the Landau-Zener tunneling as figured in the schematic diagram of Fig. \ref{fig5}(b). To clearly illustrate the above, here we consider the same parameter values as discussed in Fig. \ref{fig4} with $E_\parallel=0.1$ and the corresponding beam evolution is depicted in Fig. \ref{fig5}(c)-\ref{fig5}(f).
 
As the dispersive bands are complex valued, we observe amplification of light as shown in Fig. \ref{fig5}(c) for $E_\perp=0.01$ as $z \rightarrow \infty$. Fig. \ref{fig5}(d) shows the variation of intensity of sites with respect to propagation distance $z$ for $E_\perp=0.01$ (and other parameters as shown in Fig. \ref{fig5}(c)), where we find that the intensity of initially excited sites 0 and 1 (violet curves) first decreases up to some propagation distance $z_1$ and then increases slowly at the asymptotic limit. The value of $z_1$ depends on the system parameters, so it may become difficult to find where the asymptotic limit of site-0 and site-1 is located. However, in Fig 5(d) the value of $z_1$ is identified as $z_{1}\approx9000$. On the otherhand, due to the Landau-Zener tunneling, the given initial excitation is not confined to the site-$0$ and site-$1$ whereas the intensity at these sites in Fig. \ref{fig5}(d) deviates from zero and it increases slowly through oscillation with its neighbouring sites. Increasing $E_\perp$ to $0.05$, the dispersive bands become more complex, which results in the rapid amplification of Bloch oscillations as observed in Figs. \ref{fig5}(e) and \ref{fig5}(f). The results illustrates that due to the explicit broken nature of the $\cal{PT}$-symmetric system in the presence of $E_\perp$, amplification arises during oscillations for any parametric region. This type of amplified Bloch oscillations in non-Hermitian frequency lattices were very recently reported in \cite{amplifybo} and it has potential applications in spectrum reshaping and filtering. 
\par With these understandings on the transport dynamics of the non-Hermitian flat band lattice (1), we turn our attention towards the non-flat band case of the non-Hermitian lattice in the following section and study the associated evolution with the possibility of stable Bloch oscillations and Landau-Zener tunneling.   
\section{Dynamics in the non-flat band case}
\subsection{In the absence of electric field components}
\begin{figure}[h]
	\centering
	\includegraphics[width=1.0\linewidth]{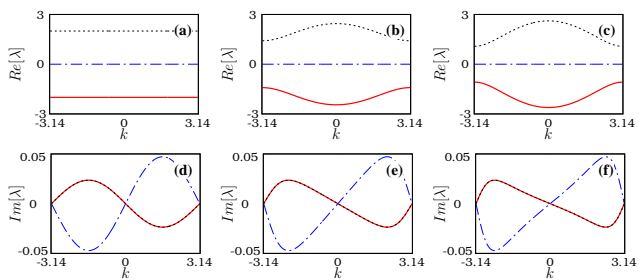}
	\caption{(Color online) (a), (b), (c)  Real parts of the band structures  and (d), (e), (f) Imaginary parts of the band structures for different phase values of the magnetic field. The imaginary parts of two of the bands exactly match with each other which are represented by a black (dotted) curve on top of the red (continuous) curves while the third band is represented by blue dash-dotted curves. (a) and (d) $\phi=\frac{\pi}{2}$, (b) and (e) $\phi=\frac{\pi}{3}$ and (c) and (f) $\phi=\frac{\pi}{4}$. Other parameters are $\gamma=0.05$ , $E_\parallel=0.0$ and $E_\perp=0.0.$}
		\label{fig6}
\end{figure}

As discussed earlier, the considered non-Hermitian lattice model supports flat band only when $\phi=l \pi$, $l=0,1$.  In other situations, all the three bands corresponding to the system are complex and dispersive.  For instance, Fig. \ref{fig6} shows the band structure in the cases $\phi=\frac{\pi}{2}$, $\frac{\pi}{3}$ and $\frac{\pi}{4}$ with $\gamma=0.05$, $E_\parallel=E_\perp=0$. 
  From Figs. \ref{fig6}(a) and \ref{fig6}(d) corresponding to the case $\phi=\frac{\pi}{2}$, it is clear that the eigenspectra are complex (where all the eigenvalues are complex) and even though the real part of the eigenvalues show no dependence on $k$, their imaginary parts depend on $k$.  In Fig. \ref{fig6}(d), the imaginary part of the band represented by the red (continuous) curve matches exactly with the one represented by black (dotted) curve where these two bands show amplifying nature for the wavevectors $-\pi<k<0$ and they show lossy nature for the wavevectors $0<k<\pi$. $Im[\lambda]$ corresponding to the band represented by blue (dashed) curve shows the opposite behaviour where it shows lossy nature for $-\pi<k<0$ and gain nature for $0<k<\pi$. Considering  the cases of $\phi=\frac{\pi}{3}$ and $\frac{\pi}{4}$, Figs. \ref{fig6}(b), \ref{fig6}(c), \ref{fig6}(e) and \ref{fig6}(f) denote the complex eigenspectra, where two of the bands show dependence on $k$ in both $Re[\lambda]$ and $Im[\lambda]$ values while the $Re[\lambda]$ of the other band does not show any dependence on $k$ where its imaginary part does show $k$ dependence.  Considering all the cases presented in Fig. \ref{fig6} it is obvious that all the bands are isolated and their complex forms indicate the spontaneously broken $\cal{PT}$-symmetric nature.  
\subsection{Super Bloch oscillations through Landau-Zener tunneling}
\par
\begin{figure}[h]
	\centering
	\includegraphics[width=1.015\linewidth]{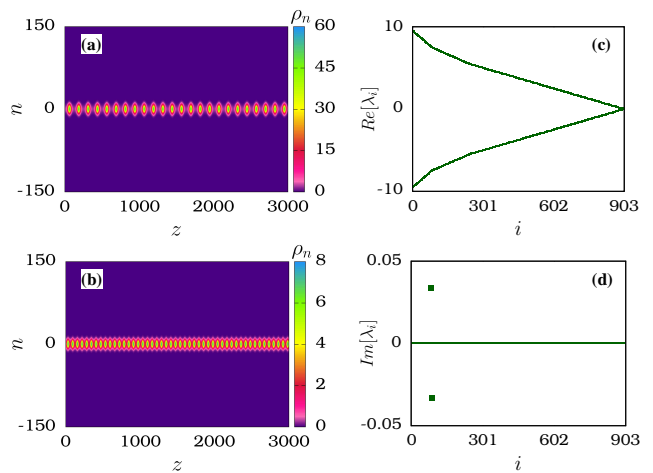}\\
	\caption{(Color online) (a) and (b) Numerical simulation using Gaussian excitation for the longitudinal electric fields $E_\parallel=0.05$ and $E_\parallel=0.1$, respectively with phase of the  magnetic field ($\phi=\frac{\pi}{2}$). (c) and (d) indicate the real and imaginary parts of the eigenvalues versus eigenvalue index by considering 903 lattice sites (i.e., 301 unit cells) corresponding to the figure (a). Other parameters are $\gamma=0.05$  and $E_\perp=0$.}
		\label{fig7}
\end{figure}

 The above discussion shows the existence of isolated bands with complex eigenspectra and we now turn our attention to studying the possibility of Landau-Zener tunneling and the associated transport dynamics in this non-flat band case. In the flat band case, even though the dispersive bands are complex, the non-dispersive flat band is real so that it can support stable localized transport for CLS initial condition in the absence of fields but this is not the case here. As all the bands are found to be complex and are isolated, there is no stable dynamics in the absence of external fields. The question of stable Bloch oscillations or stable transport dynamics in the presence of electric fields is of interest here. First, we consider the evolution of the system in the presence of $E_\parallel$ alone (that is, in the absence of $E_\perp$). In Figs. \ref{fig7}(a) and \ref{fig7}(b), we have captured the intensity evolution for the case $\phi=\frac{\pi}{2}$ and  $E_\parallel=0.05$ and $0.1$.
  To obtain the above, we have considered Gaussian type initial condition $A_{n}(0)=-C_{n}(0)=e^{-\frac{n^2}{2 \sigma^2}}$, where $\sigma=70$ and $B_{n}(0)=0$ (as considered in Sec. \ref{ptbo}), as a typical example. Figs. \ref{fig7}(a) and \ref{fig7}(b) interestingly show the existence of localized stable Bloch oscillations.  Landau-Zener tunneling induced by $E_\parallel$ makes it possible to have stable dynamics in the broken phase (showing unstable dynamics) whereas two of the eigenvalues are complex (i.e, $Re[\lambda_{82,83}]$, $Im[\lambda_{82,83}]$$\neq0$) in the presence of $E_\parallel$ as depicted in Figs. \ref{fig7}(c) and  \ref{fig7}(d). Importantly, the observed Bloch oscillations are found to be symmetric in space as well as with respect to the propagation distance $z$. Comparing Fig. \ref{fig7}(a) with Fig. \ref{fig7}(b), we observe that for weak $E_\parallel$, the intensity of Bloch oscillations is quite high and super Bloch oscillations can be observed. This type of super Bloch oscillation is realized earlier in the $\cal{PT}$ phase of the non-Hermitian tight-binding lattice with periodic forcing \cite{super}. However, such oscillations have not been found to exist even for single-site excitation in the $\cal{PT}$ broken phase of the system.  Interestingly, here we have observed super Bloch oscillations that persist for a long propagation distance $z$ with broad-site Gaussian initial excitations in the $\cal{PT}$ broken phase of the system. Strengthening $E_\parallel$ leads to a decrease in the intensity of Bloch oscillations and an increase in the frequency of Bloch oscillations.  In  the presence of $E_\parallel$, adding an additional weak transverse electric field ($E_\perp$) gives rise to amplified Bloch oscillations and further strong transverse field leads to blow-up regimes.
\section{summary}
In this paper, we have concentrated on the possibility of achieving stable Bloch oscillations in a non-Hermitian lattice model which does not support complete real eigenspectra in any of its parametric regions. Our model supports a flat band in a particular situation and we have studied the transport dynamics of the model in two different cases. 
   \par In the flat band case, we have established the following:\\
   (1) In the absence of an external electric field, the flat band case is found to have a gapless complex band structure in the region $\gamma \leq \gamma_{c}$ and has isolated bands in the region $\gamma > \gamma_{c}$.\\
  (2) The application of electric field component $E_\parallel$ induces neither amplified nor attenuated Bloch oscillations for finite values of $\gamma$ in the region of $\gamma\leq\gamma_{c}$ and unstable dynamics was observed in the region $\gamma>\gamma_{c}$ with isolated bands. Our results emphasize the existence of Bloch oscillations in the particular parametric region of the considered system.\\
   (3) With the introduction of $E_\perp$, the system is no more $\cal{PT}$-symmetric and in all the parametric regions, we have complex band structures with isolated bands. Due to this, we observe only amplifying Bloch oscillations through Landau-Zener tunneling which can be applicable in optical communications to enhance the optical signals during propagation. 
   \par In the non-flat band case, we have highlighted the following:\\
   (1) Considering the non-flat band case, complex dispersive bands are isolated and stable dynamics is not possible in the absence of external fields.\\ 
  (2) However, upon the application of $E_\parallel$, the Landau-Zener tunneling among the complex bands make super Bloch oscillations possible in the broken phase of the system.\\ 
  (3) While applying $E_\perp$ in this case, the $\cal{PT}$-symmetry of the system is explicitly broken and we observe either amplified Bloch oscillations or blow-up regimes depending on the strength of $E_\perp$. 
\par The other important aspects which we have identified in the present model is the asymmetric nature of the Bloch oscillations observed in the flat band case while the ones observed in the non-flat band case are found to be symmetric. This result may be compared with the observation made in the Hermitian flat band case \cite{flat} where asymmetric Bloch oscillations with Landau-Zener tunneling are reported (while the usual non-flat band Hermitian systems show symmetric Bloch oscillations).
\par We do believe that the observed results may open up a promising way to control light or electron transport using non-Hermitian lattices. Particularly, in contrast to Bloch oscillations observed in the gapless situation, the Bloch oscillations induced by Landau-Zener tunneling in the gapped situations are found to be of high importance and so it may be useful in the applications of optical amplification or in achieving localized transport of high-intense beam. In the future, it would be interesting to study the Bloch oscillations and Landau-Zener tunneling in non-Hermitian flat band systems along with nonlinearity \cite{disodnonlin, bononlin, dianonlin, ABnonlin}.
\section*{Acknowledgement}
J.R.P thanks the Department of Science and Technology, Government of India, for providing an INSPIRE Fellowship No. DST/INSPIRE Fellowship/2017/IF170539. The work of V.K.C forms part of the research projects sponsored by SERB-DST-MATRICS Grant No. MTR/2018/000676 and CSIR Project Grant No. 03/1444/18/EMR-II.  M.L. wishes to thank the Department of Science and Technology for the award of a SERB Distinguished Fellowship under Grant No.SB/DF/04/2017.

\end{document}